\documentstyle[prl,epsfig,aps,floats,amstex]{revtex} 

\def\ppbar{$p\overline{p}~$}             
\def\met{\mbox{${\hbox{$E$\kern-0.6em\lower-.1ex\hbox{/}}}_T\ $}} 
\def\mex{\mbox{${\hbox{$E$\kern-0.6em\lower-.1ex\hbox{/}}}_x$}} 
\def\mey{\mbox{${\hbox{$E$\kern-0.6em\lower-.1ex\hbox{/}}}_y$}} 
\def\mexy{\mbox{${\hbox{$E$\kern-0.6em\lower-.1ex\hbox{/}}}_{x,y}$}} 
\def\metsig{\mbox{$\raisebox{.3ex}{$\not$}E_T$\hspace*{0.1ex}$^{sig}\ $}}
\def\metsig{\mbox{$\raisebox{.3ex}{$\not$}E_T^{sig}\ $}}
 
\def\D0{D\O}                            
\font\eightit=cmti8
\def\r#1{\ignorespaces $^{#1}$}

\begin{document}
\lefthyphenmin=2
\righthyphenmin=3

%
%
\title{
\begin{flushright}
{\rm FERMILAB-PUB-05-009-E}
\end{flushright}
Measurement of the $W^{+}W^{-}$ Production Cross Section in $p\bar{p}$
Collisions at $\sqrt{s} = 1.96$ TeV using Dilepton Events}

\maketitle

\font\eightit=cmti8
\def\r#1{\ignorespaces $^{#1}$}
\hfilneg
\begin{sloppypar}
\noindent 
D.~Acosta,\r {16} J.~Adelman,\r {12} T.~Affolder,\r 9 T.~Akimoto,\r {54}
M.G.~Albrow,\r {15} D.~Ambrose,\r {43} S.~Amerio,\r {42}  
D.~Amidei,\r {33} A.~Anastassov,\r {50} K.~Anikeev,\r {15}
% ...
\end{sloppypar}
\vskip .026in
\begin{center}
(CDF Collaboration)
\end{center}

\vskip .026in
\begin{center}
\r 1  {\eightit Institute of Physics, Academia Sinica, Taipei, Taiwan 11529, 
Republic of China} \\
\r 2  {\eightit Argonne National Laboratory, Argonne, Illinois 60439} \\
\r 3  {\eightit Institut de Fisica d'Altes Energies, Universitat Autonoma
de Barcelona, E-08193, Bellaterra (Barcelona), Spain} \\
\r 4  {\eightit Istituto Nazionale di Fisica Nucleare, University of Bologna,
I-40127 Bologna, Italy} \\
\r 5  {\eightit Brandeis University, Waltham, Massachusetts 02254} \\
\r 6  {\eightit University of California at Davis, Davis, California  95616} \\
\r 7  {\eightit University of California at Los Angeles, Los 
Angeles, California  90024} \\
\r 8  {\eightit University of California at San Diego, La Jolla, California  92093} \\ 
\r 9  {\eightit University of California at Santa Barbara, Santa Barbara, California 
93106} \\ 
\r {10} {\eightit Instituto de Fisica de Cantabria, CSIC-University of Cantabria, 
39005 Santander, Spain} \\
\r {11} {\eightit Carnegie Mellon University, Pittsburgh, PA  15213} \\
\r {12} {\eightit Enrico Fermi Institute, University of Chicago, Chicago, 
Illinois 60637} \\
\r {13}  {\eightit Joint Institute for Nuclear Research, RU-141980 Dubna, Russia}
\\
\r {14} {\eightit Duke University, Durham, North Carolina  27708} \\
\r {15} {\eightit Fermi National Accelerator Laboratory, Batavia, Illinois 
60510} \\
\r {16} {\eightit University of Florida, Gainesville, Florida  32611} \\
\r {17} {\eightit Laboratori Nazionali di Frascati, Istituto Nazionale di Fisica
               Nucleare, I-00044 Frascati, Italy} \\
\r {18} {\eightit University of Geneva, CH-1211 Geneva 4, Switzerland} \\
\r {19} {\eightit Glasgow University, Glasgow G12 8QQ, United Kingdom}\\
\r {20} {\eightit Harvard University, Cambridge, Massachusetts 02138} \\
\r {21} {\eightit The Helsinki Group: Helsinki Institute of Physics; and Division of
High Energy Physics, Department of Physical Sciences, University of Helsinki, FIN-00044, Helsinki, Finland}\\
\r {22} {\eightit Hiroshima University, Higashi-Hiroshima 724, Japan} \\
\r {23} {\eightit University of Illinois, Urbana, Illinois 61801} \\
\r {24} {\eightit The Johns Hopkins University, Baltimore, Maryland 21218} \\
\r {25} {\eightit Institut f\"{u}r Experimentelle Kernphysik, 
Universit\"{a}t Karlsruhe, 76128 Karlsruhe, Germany} \\
\r {26} {\eightit High Energy Accelerator Research Organization (KEK), Tsukuba, 
Ibaraki 305, Japan} \\
\r {27} {\eightit Center for High Energy Physics: Kyungpook National
University, Taegu 702-701; Seoul National University, Seoul 151-742; and
SungKyunKwan University, Suwon 440-746; Korea} \\
\r {28} {\eightit Ernest Orlando Lawrence Berkeley National Laboratory, 
Berkeley, California 94720} \\
\r {29} {\eightit University of Liverpool, Liverpool L69 7ZE, United Kingdom} \\
\r {30} {\eightit University College London, London WC1E 6BT, United Kingdom} \\
\r {31} {\eightit Massachusetts Institute of Technology, Cambridge,
Massachusetts  02139} \\   
\r {32} {\eightit Institute of Particle Physics: McGill University,
Montr\'{e}al, Canada H3A~2T8; and University of Toronto, Toronto, Canada
M5S~1A7} \\
\r {33} {\eightit University of Michigan, Ann Arbor, Michigan 48109} \\
\r {34} {\eightit Michigan State University, East Lansing, Michigan  48824} \\
\r {35} {\eightit Institution for Theoretical and Experimental Physics, ITEP,
Moscow 117259, Russia} \\
\r {36} {\eightit University of New Mexico, Albuquerque, New Mexico 87131} \\
\r {37} {\eightit Northwestern University, Evanston, Illinois  60208} \\
\r {38} {\eightit The Ohio State University, Columbus, Ohio  43210} \\  
\r {39} {\eightit Okayama University, Okayama 700-8530, Japan}\\  
\r {40} {\eightit Osaka City University, Osaka 588, Japan} \\
\r {41} {\eightit University of Oxford, Oxford OX1 3RH, United Kingdom} \\
\r {42} {\eightit University of Padova, Istituto Nazionale di Fisica 
          Nucleare, Sezione di Padova-Trento, I-35131 Padova, Italy} \\
\r {43} {\eightit University of Pennsylvania, Philadelphia, 
        Pennsylvania 19104} \\   
\r {44} {\eightit Istituto Nazionale di Fisica Nucleare, University and Scuola
               Normale Superiore of Pisa, I-56100 Pisa, Italy} \\
\r {45} {\eightit University of Pittsburgh, Pittsburgh, Pennsylvania 15260} \\
\r {46} {\eightit Purdue University, West Lafayette, Indiana 47907} \\
\r {47} {\eightit University of Rochester, Rochester, New York 14627} \\
\r {48} {\eightit The Rockefeller University, New York, New York 10021} \\
\r {49} {\eightit Istituto Nazionale di Fisica Nucleare, Sezione di Roma 1,
University di Roma ``La Sapienza," I-00185 Roma, Italy}\\
\r {50} {\eightit Rutgers University, Piscataway, New Jersey 08855} \\
\r {51} {\eightit Texas A\&M University, College Station, Texas 77843} \\
\r {52} {\eightit Texas Tech University, Lubbock, Texas 79409} \\
\r {53} {\eightit Istituto Nazionale di Fisica Nucleare, University of Trieste/\
Udine, Italy} \\
\r {54} {\eightit University of Tsukuba, Tsukuba, Ibaraki 305, Japan} \\
\r {55} {\eightit Tufts University, Medford, Massachusetts 02155} \\
\r {56} {\eightit Waseda University, Tokyo 169, Japan} \\
\r {57} {\eightit Wayne State University, Detroit, Michigan  48201} \\
\r {58} {\eightit University of Wisconsin, Madison, Wisconsin 53706} \\
\r {59} {\eightit Yale University, New Haven, Connecticut 06520} \\
\end{center}

%
%
\begin{abstract}
We present a measurement of the $W^{+}W^{-}$ production cross section using 
184 ${\rm pb}^{-1}$ of \ppbar collisions at a center-of-mass energy of 1.96~TeV 
collected with the Collider Detector at Fermilab. Using the dilepton decay channel
$W^{+}W^{-}{\rightarrow}{\ell}^{+}\nu{\ell}^{-}\bar{\nu}$, where the charged 
leptons can be either electrons or muons, we find 17 candidate events compared to 
an expected background of $5.0^{+2.2}_{-0.8}$ events. The resulting $W^{+}W^{-}$ 
production cross section measurement of 
$\sigma (p\bar{p} \rightarrow W^+ W^-) = 14.6^{+5.8}_{-5.1}{\rm (stat)}\, ^{+1.8}_{-3.0} {\rm (syst)} 
\pm 0.9 {\rm (lum)}\,{\rm pb}$ agrees well with the Standard Model expectation.

\vspace{0.5cm}

\end{abstract}



\twocolumn

The measurement of the $W$ pair production cross-section in $p\bar{p}$ collisions 
at $\sqrt{s}=1.96$~${\rm TeV}$ provides an important test of the Standard Model.
Anomalous $WW\gamma$ and $WWZ$ triple gauge boson couplings~\cite{tgc}, as well as 
the decays of new particles such as Higgs bosons~\cite{higgswg}, could result in a 
rate of $W$ pair production that is larger than the Standard Model cross-section 
of $12.4 \pm 0.8\ {\rm pb}$~\cite{ww_theory}.
The first evidence for $W$ pair production was found in $p\bar{p}$ collisions by 
the CDF collaboration at $\sqrt{s}=1.8$~${\rm TeV}$~\cite{cdf1ww}. 
The properties of $W$ pair production have been extensively studied by the LEP 
collaborations in $e^+ e^-$ collisions up to $\sqrt{s}=209$~${\rm GeV}$~\cite{wwlep}, 
and have been shown to be in good agreement with the Standard Model.
The \D0 experiment has recently reported a measurement of the $W$ pair production
cross section at Run II of the Tevatron~\cite{d0}.

In this Letter we describe a measurement of the $W^{+}W^{-}$ production cross section
in the dilepton decay channel 
$W^{+}W^{-}\rightarrow \ell^{+}\nu\ell^{-}\bar{\nu}$~$(\ell = e,\mu)$,
and compare the event kinematics with Standard Model predictions. The signature 
for $W^+ W^- \rightarrow \ell^{+}\nu\ell^{-}\bar{\nu}$ events is two high-$P_T$ 
leptons and missing transverse energy, \met\!, from the undetected 
neutrinos~\cite{coords}. Jets from the hadronization of additional partons in 
the event due to initial-state radiation may be present. 
This analysis is based on $184 \pm 11\ {\rm pb^{-1}}$ of data collected by the 
upgraded Collider Detector at Fermilab (CDF) during the Tevatron Run II period.

The CDF II detector~\cite{CDF_run2} has undergone a major upgrade since the 
Run I data-taking period. The components relevant to this analysis are briefly 
described here. The Central Outer Tracker (COT) is a large-radius cylindrical 
drift chamber with 96 measurement layers organized into alternating axial and 
${\pm}2^{\circ}$ stereo superlayers \cite{COT}, and is used to reconstruct the 
trajectories (tracks) of charged particles and measure their momenta. The COT 
coverage extends to $|\eta|=1$.
A silicon microstrip detector ~\cite{SVX1,ISL} provides precise
tracking information near the beamline in the region $|\eta|<2$. 
The entire tracking volume sits inside a 1.4~${\rm T}$ magnetic field.
Segmented calorimeters, covering the pseudorapidity region $|\eta|<3.6$,
surround the tracking system.  The central ($|\eta|<1$) and forward ($1<|\eta|<3.5$)
electromagnetic calorimeters are lead-scintillator sampling devices, instrumented with proportional and 
scintillating strip detectors that measure the position and transverse profile 
of electromagnetic showers. The hadron calorimeters are iron-scintillator 
sampling detectors.
Four layers of planar drift chambers located outside the central hadron 
calorimeters (CMU) and another set behind a 60~cm thick iron shield (CMP) 
detect muons with $|\eta|<0.6$. Additional drift chambers and scintillation 
counters (CMX) detect muons in the region $0.6<|\eta|<1.0$. Gas Cherenkov 
counters~\cite{cls} measure the average number of inelastic $p\bar{p}$ collisions 
per bunch crossing and thereby determine the beam luminosity.  

A trigger system selects events with a central electron candidate with
$E_T > 18~\rm{GeV}$, a muon candidate with $P_T > 18~{\rm GeV}/c$, 
or a forward electron candidate with $E_T > 20$~GeV. For forward electrons, 
\met$>15$~GeV is also required.

Offline, electron candidates are selected in the central region by matching a well-measured track 
reconstructed in the fiducial region of the COT to an energy cluster with $E_T > 20$~GeV deposited 
in the surrounding calorimeters with identification requirements described in detail in~\cite{wzincl}.
For forward electrons ($1.2 < |\eta| < 2.0$), the track-energy cluster association utilizes a calorimeter 
seeded silicon tracking algorithm~\cite{phoenix}.

Muon candidates are selected offline by demanding $P_T > 20~{\rm GeV}/c$, energy deposition in 
the calorimeter consistent with that of a minimum ionizing particle and the same requirements 
on the reconstructed track as for central electrons. A tightly selected muon category requires 
the COT track to extrapolate to track segments in either the CMU and CMP chambers or the CMX chambers.
A loosely selected muon category requires the COT track to extrapolate to gaps in the muon chamber 
coverage.

Significant backgrounds to $W^+W^-$ production in the dilepton decay channel include Drell-Yan 
events with large \met (mismeasured in the case of $Z/\gamma^{*}\to e^{+}e^{-}, \mu^{+}\mu^{-}$ 
or due to $\nu$'s in the case of $Z/\gamma^{*}\to\tau^{+}\tau^{-}$), $W+{\rm jet}/\gamma$ events 
in which the jet or photon fakes a lepton, $t\bar{t}$ production and heavy diboson ($WZ$, $ZZ$)
production.

All lepton candidates are required to be isolated in order to suppress the background from 
fake leptons. To be isolated, the fraction of the additional $E_T$ found in a cone with radius 
$\Delta R = \sqrt{\Delta\phi^2 + \Delta \eta^2} = 0.4$ around the electron (muon) must be 
less than 10$\%$ of the electron $E_T$ (muon $P_T$). The corresponding isolation requirement 
calculated using track momenta rather than calorimeter energy is also imposed.

Candidate events are required to have two well identified, oppositely charged leptons 
(electrons or muons), and are classified as $ee$, $\mu \mu$ or $e\mu$. An event can contain 
at most one loose muon. We furthermore reject events with more than two selected leptons of 
any kind. We require events to contain no jets with $E_T > 15~{\rm GeV}$ and $|\eta|<2.5$,
where jets are reconstructed using a cone algorithm with a radius $\Delta R = 0.4$.

We require all candidate events to have \met $>25$~GeV, after the \met has been corrected for 
the escaping muon momentum when muon candidates are present. To reduce the likelihood of falsely 
reconstructed \met due to mis-measured leptons, the \met direction must have an azimuthal 
angle of at least $20^{\circ}$ from the closest lepton if the \met is less than 50 GeV.
To further reduce the Drell-Yan background, $ee$ and ${\mu}{\mu}$ candidates with a dilepton 
invariant mass in the $Z$ mass region $76 < M_{{\ell}{\ell}} < 106~{\rm GeV}/c^2$ must pass 
an additional requirement of \metsig $>$ 3\,${\rm GeV}^{1/2}$. Here, missing transverse 
energy significance is defined as \metsig~$=$~\met$/\sqrt{\Sigma E_T}$ where $\Sigma E_T$ 
is the scalar transverse energy sum over all calorimeter towers. $\Sigma E_T$ is corrected
for muons in an identical manner to the $~\met$ calculation.

The signal acceptance is computed using a large sample of 
$W^{+}W^{-}{\rightarrow}{\ell}^{+}\nu{\ell}^{-}\bar{\nu}$ events generated using the 
{\sc pythia} Monte Carlo program~\cite{pythia} and passed through a detailed detector 
simulation. $W\to\tau\nu$ decays are included and their contribution to the total 
acceptance is taken into account.
CTEQ5 parton distribution functions (PDF's)~\cite{cteq5} are used for the 
signal as well as the background Monte Carlo samples. The trigger and lepton 
identification efficiencies are measured using $Z\to\ell^{+}\ell^{-}$ data ~\cite{wzincl}.
The final acceptance estimate for $W^{+}W^{-}$ events, assuming a branching ratio 
${\rm BR}(W\rightarrow\ell\nu)=0.1068 \pm 0.0012$~\cite{PDG}, is $0.45 \pm 0.05\%$.
The number of $W^{+}W^{-}$ events expected in the dilepton decay channels is calculated using
this acceptance number and a NLO (Next-to-Leading-Order) estimate for the total $W^{+}W^{-}$ cross 
section in $p\bar{p}$ collisions at $1.96$~${\rm TeV}$ of $12.4\pm 0.8$~${\rm pb}$~\cite{ww_theory}.
This cross-section estimate uses CTEQ6 PDF's~\cite{cteq6}.

The fraction of  $W^{+}W^{-}$ events containing no reconstructed jets (``zero-jet fraction'') is 
calculated using the $W^{+}W^{-}$ {\sc pythia} Monte Carlo sample and multiplied by the ratio of 
zero-jet fractions measured in Drell-Yan data and {\sc pythia} Drell-Yan Monte Carlo. This 
scale factor, $0.96 \pm 0.06$, corrects for the underestimate of the rate of associated jet 
production by a leading-order matrix element Monte Carlo program such as {\sc pythia}.
The corrected zero-jet fraction for $W^{+}W^{-}$ events is $76 \pm 5 \%$.

The systematic uncertainty on the total acceptance for $W^{+}W^{-}$ events in the dilepton channel 
is a combination of uncertainties on the zero-jet fraction ($6\%$), choice of generator and
parton shower model ($4\%$), jet energy scale ($3\%$), lepton identification ($2\%$), trigger
efficiencies ($1\%$), modeling of the track isolation ($4\%$) and \metsig distributions ($2\%$), 
and choice of PDF ($1\%$).  We assume no correlations between these sources of uncertainty and 
combine them to give an overall $10\%$ systematic uncertainty on the $W^{+}W^{-}$ acceptance.

The Drell-Yan background 
($Z/{\gamma}^{*} \, {\rightarrow}\, e^+e^-$, $\mu^+ \mu^-$, $\tau^+ \tau^-$) 
is estimated using a combination of data and Monte Carlo samples, 
including a large sample of {\sc pythia} generated Drell-Yan events.
The Drell-Yan background estimate in the $e\mu$ channel is entirely Monte
Carlo based. The background from $Z/{\gamma}^{*} \, {\rightarrow}\, \tau^+ \tau^-$
in all detection channels is also based on Monte Carlo alone.
In the like-flavor dilepton channels $ee$ ($\mu\mu$), the background from 
$Z/{\gamma}^{*} \, {\rightarrow}\, e^+e^-$ ($Z/{\gamma}^{*} \, {\rightarrow}\, \mu^+ \mu^-$)
is estimated with a method described next that makes use of both Monte Carlo and data.
The background estimate outside the $Z$ mass window starts by counting the number of 
data events inside the $Z$ mass window that pass all the $W^{+}W^{-}$ 
selection criteria applied outside. This number of events is multiplied by 
the ratio of the number of Drell-Yan events outside and inside the $Z$ 
mass window, estimated using Monte Carlo after all out-of-window selection criteria have 
been applied. The same method is applied to estimate the background 
inside the $Z$ mass window, using the ratio of events that pass both 
\met and \metsig cuts to those that fail one or both of these 
requirements.
Monte Carlo is needed to estimate a significant contamination from 
non-Drell-Yan events in the data samples used in this procedure.
$W^{+}W^{-}$ events themselves contribute to this contamination.
This dependence of the Drell-Yan background estimate on the 
$W^{+}W^{-}$ cross-section, and vice-versa, is resolved
by iteratively finding a common solution to both. Statistical
uncertainties on the data dominate the final systematic uncertainty on the 
Drell-Yan background.

We estimate the fake lepton background contribution by applying a
$P_T$ dependent lepton fake rate to $\ell\,+\, \met +\,d$ events, 
where $d$ denotes any object which could fake a lepton.
Such events must pass all other $W^+W^-$ selection criteria.
The lepton fake rates are defined by the ratio $N_{\ell}/N_{d}$.
Fakeable objects counted in the denominator ($N_{d}$) are jets with $E_T>20$ GeV 
and $|\eta|<2$ for electrons and tracks with $P_{T}>20~{\rm GeV}/c$ and
$E/P<1$ for muons.
The numerator ($N_{\ell}$) is the number of objects passing all lepton
identification and isolation criteria. The lepton fake rates are
determined using large samples of jet triggered data with jet $E_{T}$ 
thresholds in the range $20-100$~${\rm GeV}$, correcting for the presence 
of real leptons from $W$ and $Z$ production. The probability for an object 
to fake a lepton is of the order $10^{-4}$ to $10^{-3}$ depending on the 
lepton type and detector region. Studies of the variations of fake rates 
between jet samples with different trigger thresholds and using various 
definitions of the fakeable objects have been performed. The estimated 
systematic uncertainty on this background is 40$\%$.

The $W\gamma$ background estimate is derived using a leading-order Monte Carlo 
generator for the process 
$p\bar{p}\to W\gamma X\to \ell\nu\gamma X$~\cite{prd41_1990_1476}, 
which has been interfaced to {\sc pythia} for the purposes of parton 
showering and hadronization. The sample is normalized to a NLO calculation 
of the $W\gamma$ cross section~\cite{wg_nlo}. The $W\gamma$ background that 
is double-counted in the fake lepton background estimate described above is 
determined to be negligible. 

The remaining backgrounds from $t\bar{t}$, $WZ$ and $ZZ$ production
are calculated using Monte Carlo samples generated with {\sc pythia} and 
normalized to NLO cross-sections. 
The background from top pair production in the dilepton
decay channel ($t\bar{t}\,{\rightarrow}\,W^+ b W^- \bar{b}\,{\rightarrow}\,\ell^+ \nu b \ell^- \bar{\nu} \bar{b}$)
is greatly reduced by the zero jet requirement.
The background from $WZ$ production has two main contributions: 
$WZ\,{\rightarrow}\,q\bar{q}' \ell^+ \ell^-$, which is largely rejected by the zero jet requirement,
and $WZ\,{\rightarrow}\,{\ell}{\nu}{\ell}^{+}{\ell}^{-}$, which is largely rejected by the veto on 
trilepton events. The background coming from $ZZ$ production is predominantly due to 
$ZZ\,{\rightarrow}\,{\ell}^{+}{\ell}^{-}{\nu}\bar{\nu}$.
The final systematic uncertainty on the total background estimate is approximately 
$45\%$, dominated by the systematic uncertainty on the Drell-Yan background.

\begin{figure}[!ht]
 \begin{center}
   \mbox{\epsfig{file=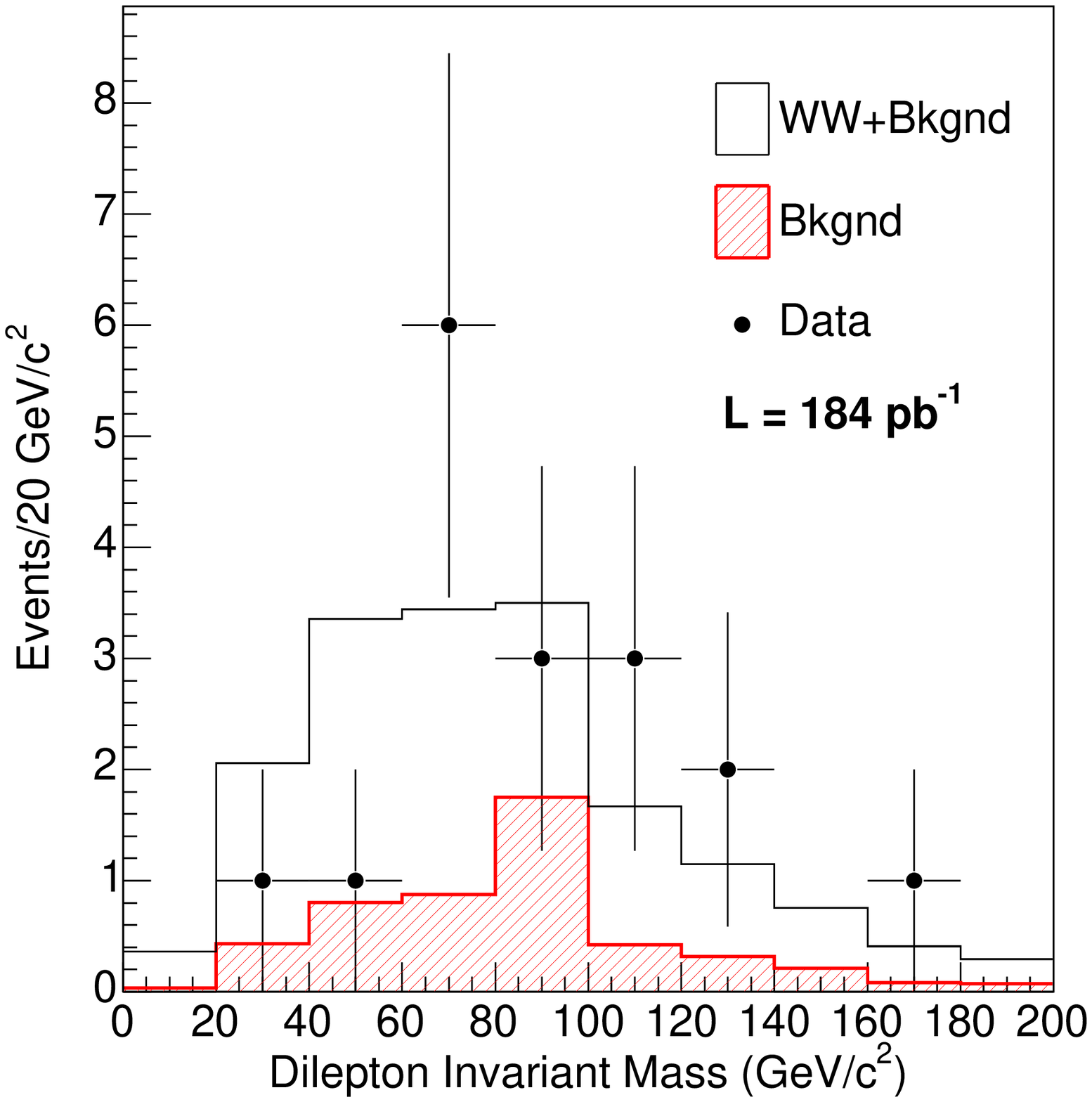,width=0.95\columnwidth,clip=}}
   \mbox{\epsfig{file=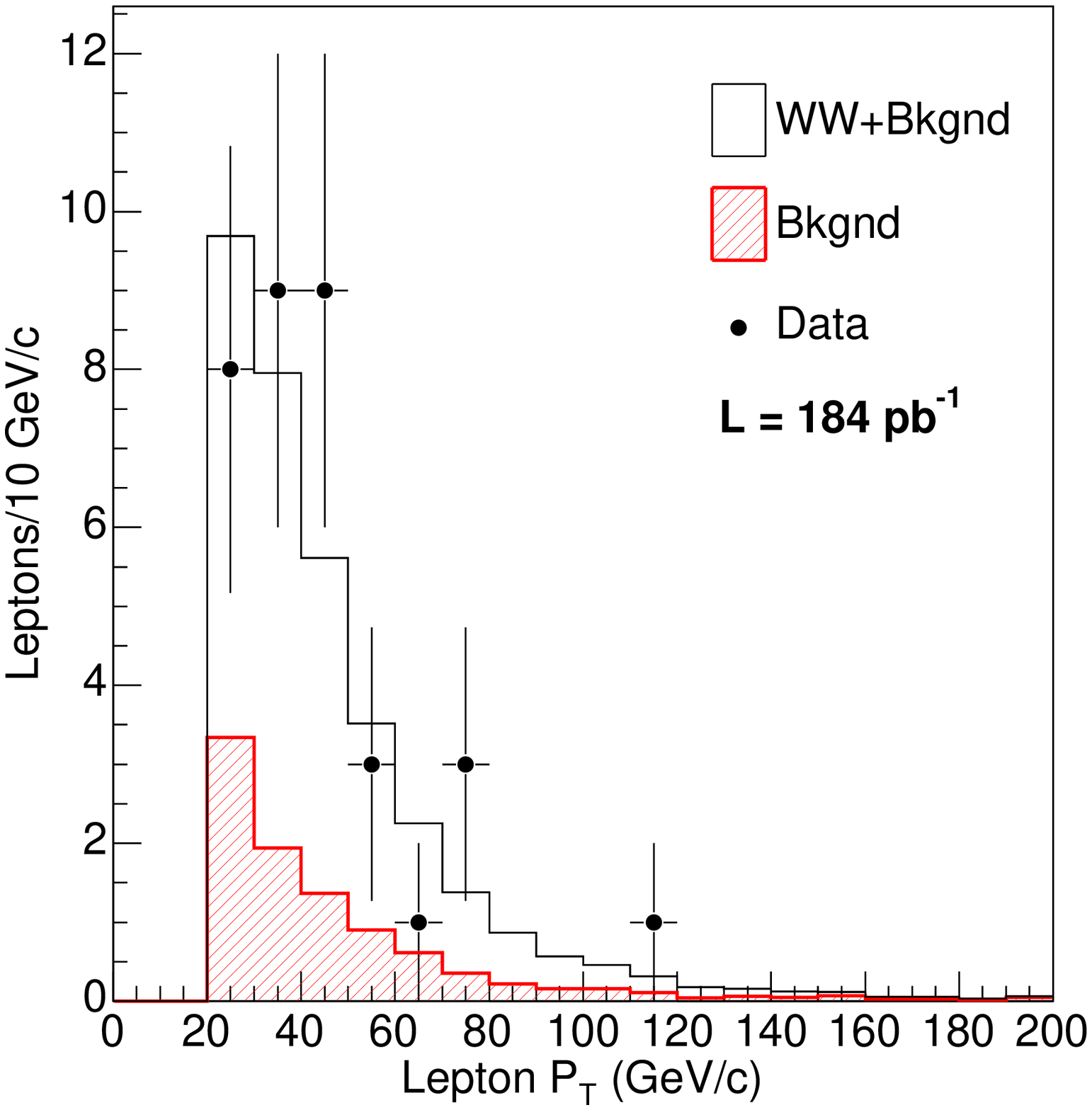,width=0.95\columnwidth,clip=}}
 \end{center}
\caption{The dilepton mass (top) and lepton transverse momentum distribution
(bottom) for the candidate events in comparison with the Standard Model 
expectation. Kolmogorov-Smirnov tests of these distributions yield $p$-values 
of 13$\%$ (top) and 78$\%$ (bottom).}
\label{fig:ww_kine}
\end{figure}

The signal and background expectations are summarised in Table~\ref{table:grand_summary}, 
together with the number of data events passing the selection criteria~\cite{wz_zz_overlap}.
The measured cross section is~:
\begin{center}
$\sigma(p\bar{p}\rightarrow W^+ W^- ) = 14.6^{+5.8}_{-5.1} {\rm (stat)} ^{+1.8}_{-3.0} {\rm (syst)} \pm 0.9 {\rm (lum)}$~${\rm pb}$
\end{center}
where the systematic uncertainty is a combination of the uncertainties on the signal
acceptance and background estimates. The third uncertainty corresponds to a $6\%$ uncertainty from 
the integrated luminosity measurement. The dilepton mass and lepton transverse momenta distributions
are shown in Figure~\ref{fig:ww_kine}. There is no evidence for statistically significant discrepancies 
in either the dilepton mass or lepton transverse momentum distributions, which could indicate the 
presence of poorly estimated backgrounds or physics beyond the Standard Model.
\begin{table}[htbp]
 \renewcommand{\arraystretch}{1.1}
 \centering 
 \begin{tabular}{l||c|c|c}
  & $ee$ & ${\mu}{\mu}$ & $e{\mu}$  \\ \hline
 $Z/{\gamma}^{*}{\rightarrow}\ell^{+}\ell^{-}$ & $0.21_{-0.16}^{+1.29}$ & $0.43_{-0.38}^{+1.56}$ & 0.43 $\pm$ 0.14 \\
 $WZ$  &  0.29 $\pm$  0.03 &  0.33 $\pm$  0.03 &  0.15 $\pm$  0.02  \\
 $ZZ$  &  0.35 $\pm$ 0.04 &  0.34 $\pm$  0.04 &  0.011 $\pm$  0.002  \\
 $W+\gamma$  &  0.48 $\pm$  0.13 &  -  &  0.57 $\pm$  0.13  \\
 $t\bar{t}$  &  0.021 $\pm$  0.011 &  0.012 $\pm$  0.007 &  0.046 $\pm$  0.018  \\
 Fake  &  0.52 $\pm$  0.19 &  0.17 $\pm$  0.16 &  0.65 $\pm$  0.37  \\ \hline
 Background  &  $1.9_{-0.3}^{+1.3}$ &  $1.3_{-0.4}^{+1.6}$ &  1.9 $\pm$  0.4 \\ \hline
 $W^{+}W^{-}$ Signal  &  2.6 $\pm$  0.3 &  2.5 $\pm$  0.3 & 5.1 $\pm$  0.6  \\ \hline
 Expected  &  $4.5_{-0.5}^{+1.4}$ &  $3.8_{-0.5}^{+1.6}$ &  7.0 $\pm$  0.8  \\ \hline
 Observed  & 6 & 6 & 5  \\ 
\end{tabular}
 \caption{Estimated backgrounds, expected $W^{+}W^{-}$ signal and the observed 
number of events in $184\,{\rm pb^{-1}}$ for each dilepton category. 
The signal expectation assumes a total $W^{+}W^{-}$ cross section of $12.4$~${\rm pb}$.
Systematic uncertainties are included.}
\label{table:grand_summary}
 \end{table}

We have performed an alternative measurement of the $W^+ W^-$ production
cross section, which tests the robustness of our result in a sample with different 
signal and background composition.  The event selection is based on the ``lepton+track''
analysis used for our measurement of the $t\bar{t}$ production cross section in the 
dilepton channel~\cite{topdil}. 

There are two important differences between the lepton+track analysis and our main analysis.
Firstly, one of the two lepton candidates is only required to be an isolated track.
Secondly, all events must pass a $\metsig$ requirement of 
$\metsig > 5.5~{\rm GeV}^{1/2}$ where here, the $E_T$ sum 
is made over all jets with $E_T > 5~{\rm GeV}$. 
The candidate isolated track must have $P_T > 20\ {\rm GeV}/c$ 
and be in the range $|\eta|< 1$. Again, only events with no jets 
are considered.  The overall acceptance is $0.42 \pm 0.05$\%, 
similar to the acceptance for the main analysis. The increased
acceptance for dilepton events where electrons or muons pass 
through gaps in the calorimetry or muon system and for single
prong hadronic decays of the $\tau$ lepton from $W\rightarrow\tau\nu$
is offset by the more restrictive $\metsig$ cut required to
control the Drell-Yan background.

The numbers of observed events, the expected Standard Model backgrounds and the
predicted $W^{+}W^{-}$ signal are compared for both analyses in Table~\ref{ltrack_summary}.
The higher background rates for the lepton+track analysis are mainly due to the fake 
lepton background contribution coming from the isolated track.
The resulting cross section measurement using the lepton+track selection is:  
\begin{center}
$\sigma(p\bar p \rightarrow W^+ W^-) = 24.2 \pm 6.9 (\rm{stat})\, ^{+5.2}_{-5.7}(\rm{syst})\pm 1.5\mathrm{(lum)}$~pb.
\end{center}

The two measurements are statistically compatible with one another given an estimated
43\% overlap in signal acceptance. Since combining the results of these two analyses 
does not result in a significant reduction of the uncertainty, we quote as the 
final result of this measurement the analysis with the best $a\ priori$ sensitivity, 
which is the analysis summarized in Table I.

\begin{table}[htbp]
 \renewcommand{\arraystretch}{1.1}
 \centering
  \begin{tabular}{l||c|c}
& MAIN & LTRK \\ \hline
  Drell-Yan ($Z/{\gamma}^{*}{\rightarrow}\ell^{+}\ell^{-}$)  &  $1.06^{+2.03}_{-0.44}$  &  $1.81_{-1.38}^{+2.36}$  \\
  $WZ$                                                     &  0.76$\pm$0.06           &  $1.01 \pm  0.24$        \\
  $ZZ$                                                     &  0.70$\pm$0.07           &  $0.76 \pm  0.18$        \\
  $W+\gamma$                                               &  1.06$\pm$0.19           &  $0.33 \pm  0.13$        \\
  $t\bar t$                                                &  0.078$\pm$0.023         &  $0.18 \pm  0.04$        \\
  Fake                                                     &  1.34$\pm$0.66           &  $7.96 \pm  3.47$        \\ \hline
  Background                                               &  5.0 $_{-0.8}^{+2.2}$ &  $12.1_{-3.8}^{+4.2}$    \\ \hline
  $W^{+}W^{-}$ Signal                                      &  10.20 $\pm$  1.19       &  $10.23 \pm  1.37$       \\ \hline
  Expected                                                 &  $15.2_{-1.5}^{+2.5}$    &  $23.0_{-4.0}^{+4.4}$    \\ \hline
  Observed                                                 &  17                      &  $32$                    \\
  \end{tabular}
\caption{Estimated backgrounds, expected $W^{+}W^{-}$ signal and the observed number of 
events for both the main (MAIN) analysis using $184\,{\rm pb^{-1}}$ 
and the lepton+track (LTRK) analysis using $197\,{\rm pb^{-1}}$. 
The signal expectation assumes a total $W^{+}W^{-}$ cross section of $12.4$~${\rm pb}$.
Systematic uncertainties are included.}
\label{ltrack_summary}
\end{table}

In summary, we have measured the $W^{+}W^{-}$ cross-section in 
$p\bar{p}$ collisions at $\sqrt{s}=1.96$~${\rm TeV}$ to be
$14.6^{+6.1}_{-6.0}~{\rm pb}$. This is based on the observation of 17 
events consistent with originating from $W$ pair production
and subsequent decay to two charged leptons, compared to a total estimated
background of $5.0^{+2.2}_{-0.8}$ events. The measured cross section
is consistent with a NLO Standard Model prediction and is 
corroborated by an independent lepton + track analysis.

\begin{center}
\textbf{Acknowledgments}
\end{center}

We thank the Fermilab staff and the technical staffs of the participating
institutions for their vital contributions. We also thank
John~Campbell and Keith~Ellis for many useful discussions.
This work was supported by the
U.S. Department of Energy and National Science Foundation; the Italian
Istituto Nazionale di Fisica Nucleare; the Ministry of Education, Culture,
Sports, Science and Technology of Japan; the Natural Sciences and
Engineering Research Council of Canada; the National Science Council of the
Republic of China; the Swiss National Science Foundation; the A.P. Sloan
Foundation; the Bundesministerium fuer Bildung und Forschung, Germany; the
Korean Science and Engineering Foundation and the Korean Research
Foundation; the Particle Physics and Astronomy Research Council and the
Royal Society, UK; the Russian Foundation for Basic Research; the Comisi\'on
Interministerial de Ciencia y Tecnolog\'{\i}a, Spain; and in part by the European
Community's Human Potential Programme under contract HPRN-CT-2002-00292, Probe
for New Physics.


\begin{thebibliography}{99}
\bibitem{coords}
  We use a coordinate system where $\theta$\ is the polar angle to the
  proton beam, $\phi$\ is the azimuthal angle about this beam axis,
  and $\eta$ is the pseudorapidity defined as $-\ln(\tan(\theta/2))$.
  The transverse momentum of a particle is denoted by $P_T \equiv P
  \sin\theta$. The analogous quantity defined using energies, defined
  as $E_T\equiv E \sin \theta$, is called transverse energy.
  Missing transverse energy, \met, is defined as the magnitude of
  $-\sum_i E_T^i \hat{n}_i$, where $\hat{n}_i$\ is a unit vector in
  the azimuthal plane that points from the beamline to the $i^{\rm th}$
  calorimeter tower.
\end{thebibliography}
\end{document}